\documentclass[tradiabstract]{aa}
\usepackage{txfonts}
\usepackage{graphicx}
\usepackage{natbib}
\usepackage{longtable}
\bibpunct{(}{)}{;}{a}{}{,} 

\newcommand{\HI}{HI}
\newcommand{\HII}{HII}
\newcommand{\Htwo}{$\rm{H_2}$}
\newcommand{\Gnaught}{$G_0$}
\newcommand{\HIunits}{$10^{20}$ cm$^{-2}$}
\newcommand{\UVunits}{$10^{-15}$ ergs cm$^{-2}$ s$^{-1}$ \AA$^{-1}$}
\newcommand{\dtg}{$\delta/\delta_0$}
\newcommand{\correction}[1]{#1} 

\begin{document}

\title{The volume densities of giant molecular clouds in M83}

\author{Jonathan S. Heiner\inst{1,2}
  \and Ronald J. Allen\inst{2} 
  \and O. Ivy Wong\inst{3}
  \and Pieter C. van der Kruit\inst{1}}

\offprints{J.S. Heiner, \email{heiner@astro.rug.nl}}

\institute{Kapteyn Astronomical Institute, University of Groningen, PO Box 800, 9700 AV Groningen, the Netherlands
  \and Space Telescope Science Institute, Baltimore, MD 21218, USA
  \and Astronomy Department, Yale University, PO Box 208101, New Haven, CT 06520-8101, USA}

\date{Received date June 20, 2008 / Accepted date July 24, 2008}

\abstract{Using observed GALEX far-ultraviolet (FUV) fluxes and VLA images of the 21-cm \HI\ column densities, along with estimates of the local dust abundances, we measure the volume densities of a sample of actively star-forming giant molecular clouds (GMCs) in the nearby spiral galaxy M83 on a typical resolution scale of 170 pc.\\
Our approach is based on an equilibrium model for the cycle of molecular hydrogen formation on dust grains and photodissociation under the influence of the FUV radiation on the cloud surfaces of GMCs.\\
We find a range of total volume densities on the surface of GMCs in M83, namely 0.1 - 400 cm$^{-3}$ inside $R_{25}$, 0.5 - 50 cm$^{-3}$ outside $R_{25}$. Our data include a number of GMCs in the \HI\ ring surrounding this galaxy. Finally, we discuss the effects of observational selection, which may bias our results.}


\maketitle

\section{Introduction}

This paper aims to measure the total gas densities in a sample of giant molecular clouds (GMCs) in the nearby spiral M83 (NGC 5236), using a method initially proposed by \citet{1997ApJ...487..171A}. This method is based on the simple (but unavoidable) fact that ultraviolet photons from young, newly-formed stars will react back on the surrounding parent GMCs, dissociating the molecular gas on their surfaces and turning the (virtually invisible) \Htwo\ into its easily-detected atomic form. The motivation for this approach was provided by the discovery of \citet{1986Natur.319..296A} that the \HI\ delineating the spiral arms in a nearby galaxy showed a large-scale morphology that was more consistent with photodissociation near the \HII\ regions than it was with compression of the \HI\ farther upstream in the spiral shock. The presence of atomic hydrogen is then indicative of a photodissociation region (PDR). The method was first applied in some detail to M101 by \citet{2000ApJ...538..608S} and more recently to M81 by \citet{2008ApJ...673..798H}.

The large, nearby spiral M83 has been a frequent target of searches for molecular gas using the CO(1-0) spectral line, which is easily detectable in M83. However, such studies have only recently begun to be carried out with sufficient linear (spatial) resolution to discern differences in the location of molecular gas and the emissions from young, hot stars.
For example, \citet{1991ApJ...381..130L} describe the presence of molecular (CO) emission ~300 pc downstream from M83's eastern spiral arm dust lane, detected with a $5\arcsec\ \times 12\arcsec$ beam.
\citet{1999ApJ...513..720R} find that the CO emission is spatially separated from the dust lane as well as the young stars on scales of a few hundred parsec. They suggest UV heating, cosmic-ray heating or a two-component molecular phase to explain this morphology. Their observations show features that show similarity to the largest GMCs in the Milky Way with masses on the order of several millions of solar masses.
\citet{1993ApJ...414...98W} explore the CO content in the central kiloparsec of M83 and infer a low-density component ($n_{H_2} \lesssim 10^3-10^4 \rm{cm}^{-3}$) and a warm (above 50 K), high-density component ($n_{H_2} \gtrsim 10^4-10^5 \rm{cm}^{-3}$).  
\citet{2002AJ....123.1892C} study CO(1-0), CO(2-1) and neutral gas in M83, observing a strong truncation of the molecular disk at 6\arcmin\ accompanied by a warped atomic outer disk. Using the conventional assumptions about how to convert CO surface brightness to \Htwo\ column density, they conclude that roughly 80\% of the total gas mass in M83 is \Htwo. \correction{The} \HI\ 21 cm and CO surface brightness are found to be correlated, but with a complex pattern of offsets. They speculate that the temperature of the CO gas is $>$ 20K in the nucleus and $<$ 7K in the outer disk. 
\citet{2004A&A...413..505L} give a full overview of previous observations and present full CO (1-0) and CO (2-1) maps based on thousands of telescope pointings, but with a modest spatial resolution of $\sim 1$ kpc. They conclude that the molecular gas spiral arms mostly trace the dust lanes. They expect the \Htwo\ mass to dominate that of \HI\ within 7.3 kpc of the center. At their linear resolution, CO and \HI\ emissions are correlated strongly within the optical disk. 
\citet{2005A&A...441..491V} look into various tracers of star formation, including Polycyclic Aromatic Hydrocarbon (PAH) emission lines, another potential indicator of PDRs. Various combinations of these lines prove to be good tracers of star forming regions in M83. They are found predominantly in the spiral arms. We studied the occurrence of PAHs near PDRs in M81 \citep{2008ApJ...673..798H} previously and confirmed that they are found near the star forming regions in almost all cases. Similar M83 data was not available at this time.

\begin{figure*}[t!]
  \centering
  \includegraphics[width=17cm]{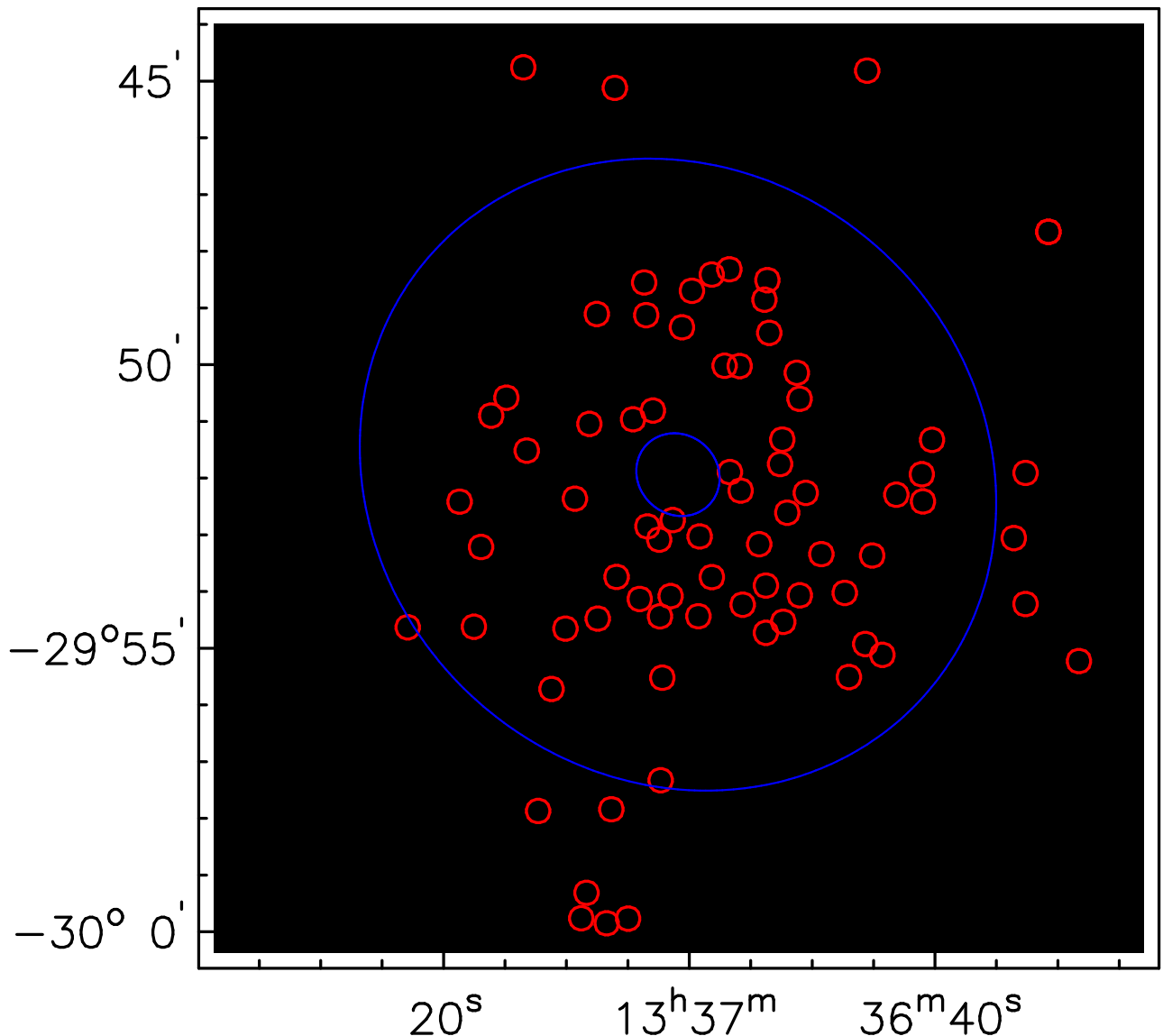}
  \caption{The location of the candidate PDRs are shown with the GALEX FUV image in the background. Not all FUV sources are visible on this image, but all sources are sources of FUV radiation. The inner circle signifies a 1 kpc galactocentric radius, the outer one signifies $R_{25}$.}
  \label{fig:locplot}
\end{figure*}

\citet{2005ApJ...619L..79T} find and discuss sites of recent star formation in the extreme outer disk of M83, associated with the warped \HI\ disk of M83, which raises questions about star formation efficiency and modes of star formation. The nature of this outer disk is investigated further in \citet{2007ApJ...661..115G}, who find evidence that individual young stars are responsible for the UV emission in the outer regions. \citet{2007AJ....134..135Z} find that these UV sources are quite common out to $2 \times R_{25}$. In this paper we include candidate PDRs outside the main optical disk of M83 in an effort to probe the amount of available gas in this environment.

The suspected high densities of molecular gas in M83, the morphology of the gas in the spiral arms, and the evidence of recent star formation in the outer regions, make this galaxy an excellent target for a study of photodissociated atomic hydrogen and the volume densities of M83's GMCs using our method.

In \S \ref{sec:method}, we explain our approach and the data we used. In \S \ref{sec:results} we present our results. We discuss and summarize these results in \S \ref{sec:discussionconlusions}.

\section{Method}
\label{sec:method}

In this paragraph we briefly describe the data and how we used it. The same method that we used in \citet{2008ApJ...673..798H} has been applied here. For every candidate PDR, the far-UV flux, local \HI\ column densities, dust-to-gas ratio and separation between the UV source and the surrounding \HI\ features were determined to calculate the total hydrogen volume density. 

The distance to M83 is taken to be 4.5 Mpc, as measured from Cepheids by \citet{2003ApJ...590..256T}. We used GALEX UV data of M83, where the FUV and near-UV (NUV) images were matched for astrometric accuracy. The images have an estimated angular resolution of 8\arcsec, which is a bit poorer than the generic GALEX resolution, since M83 is on the edge of the GALEX field of view due to pointing constraints. We did not use the NUV image otherwise. We obtained the 21 cm \HI\ data from the THINGS team (Walter et al., 2008, \aj, submitted). The radio beam size is 10.4 $\times$ 5.6\arcsec. The data were analyzed using the Groningen Image Processing System GIPSY \citep{1973A&A....27...77E,1992ASPC...25..131V,2001ASPC..238..358V}.

Although spatial completeness was not our aim, we attempted to select our candidate PDRs \correction{to be bright and isolated on the GALEX image and} to have a uniform spread in galactocentric radius as well as FUV flux\correction{, in a similar way as was done in \citet{2008ApJ...673..798H}.} First, GIPSY's POINTS routine was used to automatically detect FUV sources with a signal-to-noise of 20 or more. \correction{The other input parameters were kept at their default value. This yielded 86 sources\footnote{This translates into a typical rms noise in the FUV flux of 1\%.}.} Using a histogram of the distribution of the sources with galactocentric radius, we randomly removed a small number of sources to even out the distribution. \correction{We also removed sources that matched detections in the HST Guide Star Catalog v2.2, see e.g. \citet{2008AJ....136..735L}.} We manually added some sources to improve the spread in galactocentric radius, selected by eye. \correction{These sources stood out visually, but we did not enforce a specific signal-to-noise ratio.} Finally, these sources were automatically fitted with gaussians. In some two or three cases this led to visible confusion from nearby FUV sources as can be seen in the location plot (Fig. \ref{fig:locplot}). \correction{The final source count was 76. These sources are listed in Table \ref{tab:locations}.}

\begin{figure}[tb!]
  \resizebox{\hsize}{!}{\includegraphics{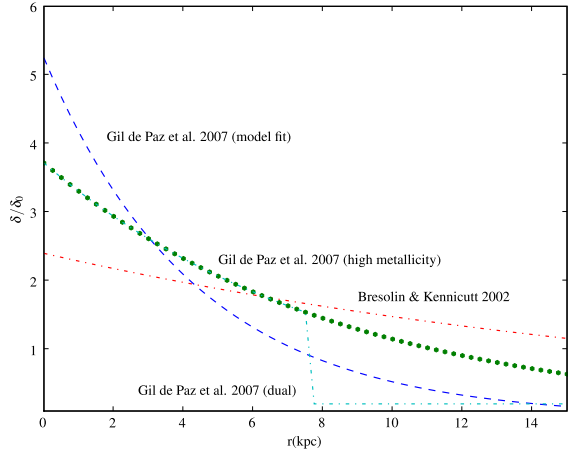}}
  \caption{Different possible dust-to-gas ratio models are plotted. There are three models suggested by \citet{2007ApJ...661..115G} and Gil de Paz (private communication, 2008), namely their high metallicity fit (our default one), their fit based on PDR modeling and a slope based on the high metallicity fit inside $R_{25}$ combined with a constant (low) metallicity in the outer regions. We also considered the one based on \citet{2002ApJ...572..838B}.}
  \label{fig:dust}
\end{figure}

We used the tilted ring model from \citet{1974ApJ...193..309R} to correct for a different position angle and inclination at larger radii. This deprojection correction changes the actual galactocentric radii of the candidate PDRs as well as the assumed deprojected separation between the UV source and its associated \HI\ patches. This information was used in combination with the metallicity data from \citet{2007ApJ...661..115G} to obtain the dust-to-gas ratio \citep[see][]{1990A&A...236..237I}. We used
\begin{equation}
  \log{\delta/\delta_0} = -0.051 R + 0.43,
  \label{eqn:GP07h}
\end{equation}
where $\delta/\delta_0$ is the local dust-to-gas ratio scaled to the value in the solar neighborhood. We adopt a solar metallicity of 8.69 \citep{2001ApJ...556L..63A}. The galactocentric radius $R$ is in kiloparsec. In subsequent plots we normalize $R$ with $R_{25}$, which is 7.63 kpc (or $\approx 6\arcmin$) from \citet{1992MNRAS.259..121V}. We assume that this relation covers the full range of galactocentric radii we investigated. It becomes clear from the literature that the metallicity in the outer regions of M83 is still uncertain and results have only recently become available. For this reason we explore a number of alternatives. We also considered, among others, \citet{2002ApJ...572..838B}. The different alternatives are plotted in Fig. \ref{fig:dust}. \correction{We have adopted the most recent high metallicity fit in \citet{2007ApJ...661..115G} as our preferred model.}

\begin{figure*}[tb!]
  \centering
  \includegraphics[width=8cm]{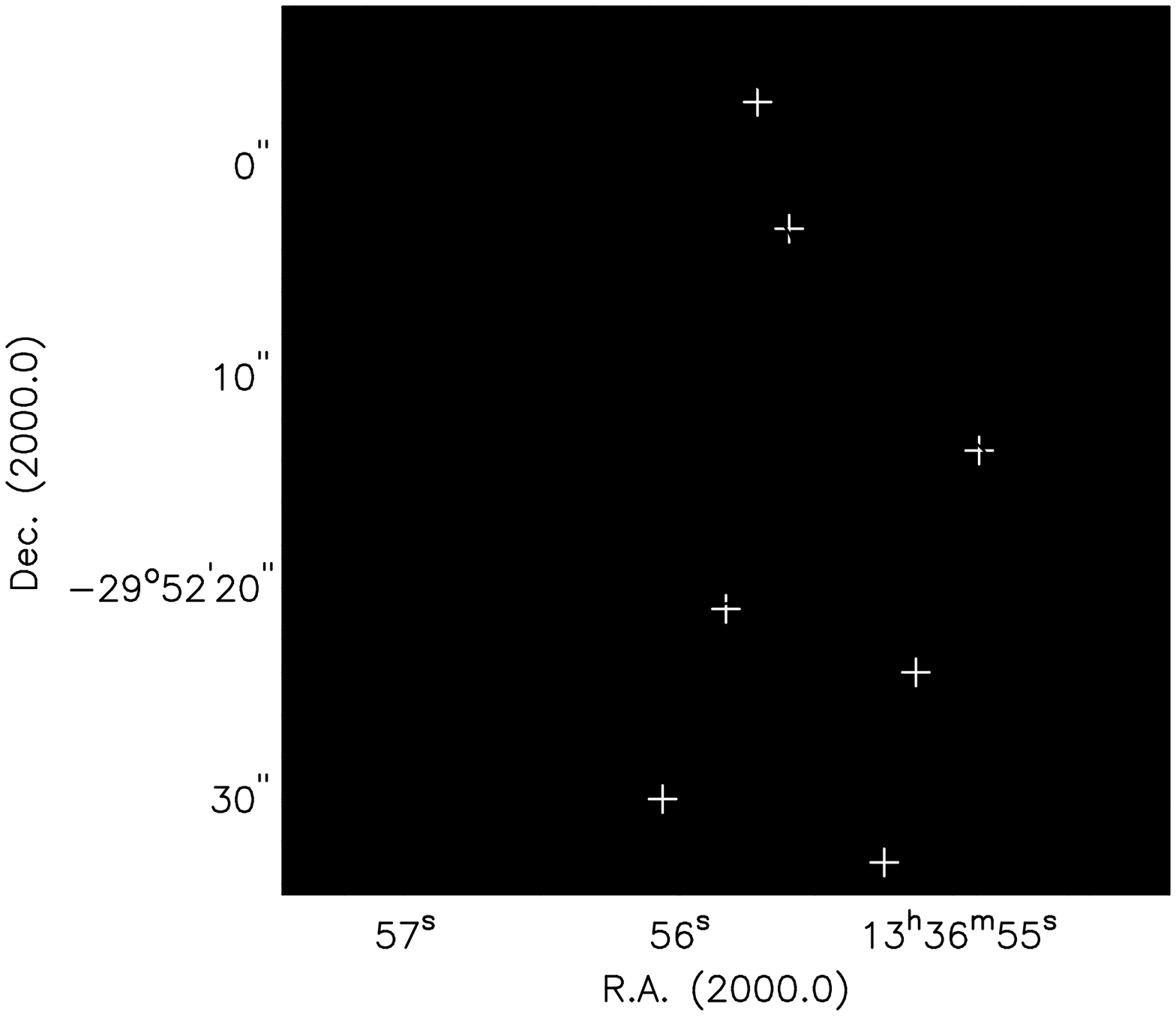}
  \includegraphics[width=8cm]{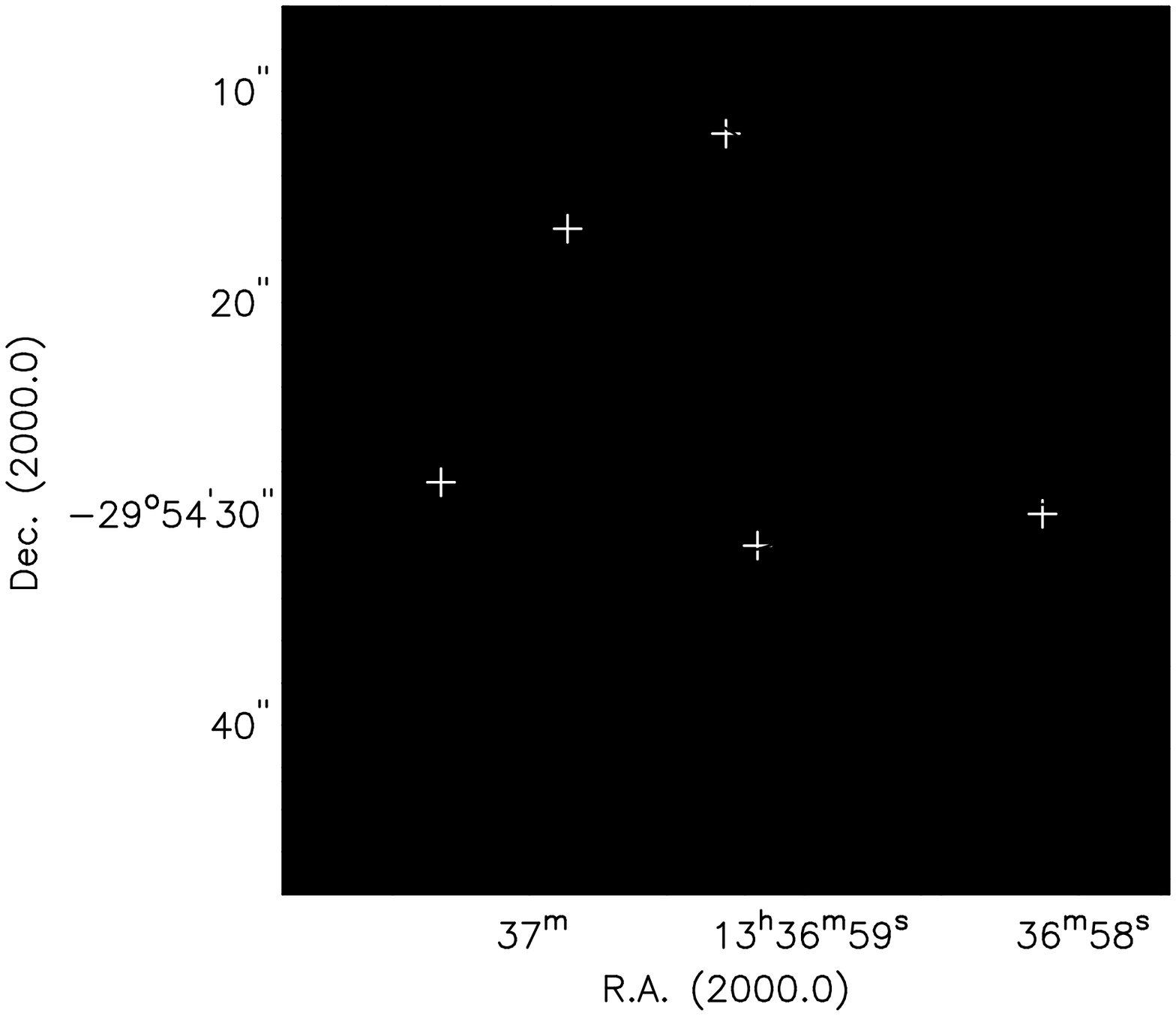}
  \caption{Sources 4 (left panel) and 24 (right panel) are plotted as examples of candidate PDRs, showing \HI\ grayscales and FUV contours. The location of the central source is marked with a black star, the locations of the \HI\ patches are marked with white crosses. The FUV contours start at $7 \times 10^{-19} \rm{ergs~cm^{-2}~s^{-1}}$ \AA$^{-1}$ and each consecutive level is 1.34 times the previous one. The \HI\ grayscale levels are constructed similarly, see Table \ref{tab:locations}. Both sources show a central source surrounded  by \HI\ complexes, where the distribution is smoother in Source 24. The full set of sources is available in the electronic version of this paper (without the white crosses).}
  \label{fig:examples}
\end{figure*}

The (local background subtracted) FUV fluxes of the candidate PDRs were determined next. For each such region, the distances from the central source to the nearest \HI\ patches was determined. An \HI\ patch is defined as a local maximum in the \HI\ column density. A background level of $2 \times 10^{20}~\rm{cm}^{-2}$ was subtracted within $0.9 R_{25}$ and $5 \times 10^{19}~\rm{cm}^{-2}$ outside this radius.
\correction{These values are based on the general background level of \HI\ emission surrounding M83. We also measured \HI\ background levels within the optical disk of M83 as well as considered minimum column density values throughout M83.}
This is an attempt to isolate the \HI\ formed by dissociating radiation. Since the background level is close to the sensitivity limit, some measurements end up close to zero. 
\correction{The resulting values of $n$ are generally not very sensitive to varying the background \HI\ column. Rewriting Equation \ref{eqn:n} to yield $n$ (see Equation 4 in \citet{2008ApJ...673..798H}) shows that $n$ does not change much as long as the subtracted background column remains substantially \correction{lower} than the scaling factor of $7.8 \times 10^{20}~\rm{cm}^{-2}$, as is the case here.}
We measured distances out to ~22\arcsec. Those distances were then used together with the FUV flux to calculate the incident flux impinging on each \HI\ patch. A 1500\AA\ foreground extinction correction of 0.52 mag based on \citet{1998ApJ...500..525S} and \citet{2007ApJS..173..185G} was applied to the FUV fluxes. Two example sources are plotted in Fig. \ref{fig:examples}.

Finally, all these elements were used to calculate the total hydrogen volume density by inverting Equation A2 from \citet{2004ApJ...608..314A} and including a variable dust-to-gas ratio \citep[cf. Equation 6 in][]{2004ASSL..319..731A}. The final equation to be inverted is:
\begin{equation}
  N_{HI} = \frac{7.8 \times 10^{20}}{\delta/\delta_0} \ln\left[1+\frac{106G_0}{n}\left(\frac{\delta}{\delta_0}\right)^{-1/2}\right]~\rm{cm^{-2}}
  \label{eqn:n}
\end{equation}
where $N_{HI}$ is the (background subtracted) atomic hydrogen column density (in $\rm{cm^{-2}}$), \dtg\ is the dust-to-gas ratio scaled to the solar neighborhood value, \Gnaught\ is the incident flux measured at the \HI\ patch (\correction{the same \Gnaught\ as used in e.g. \citet{2004ApJ...608..314A}, Appendix B -} see \citet{2008ApJ...673..798H} for a more extended description of how this is measured) and $n$ is the total hydrogen volume density, where $n = n_{HI} + 2 n_{H_2}\ \rm{(cm^{-3})}$. This gas is mostly atomic on the surface of the GMC and mostly molecular deep inside the cloud.

\section{Results}
\label{sec:results}

The measured FUV fluxes are presented first, followed by the associated \HI\ column densities. The derived incident fluxes \Gnaught\ are presented next, followed by the resulting total hydrogen volume densities. Finally, we deal with a number of selection effects.

The locations of our candidate PDRs are shown in Fig. \ref{fig:locplot} and tabulated in Table \ref{tab:locations}. We attempted to get a selection of sources representative in both their galactocentric radius as well as their UV flux, as was mentioned previously. The table also includes the aperture of each FUV source within which its flux was determined, the starting grayscale level in the detailed plots and the grayscale level increment, where each subsequent level is the previous level multiplied by the increment. \correction{While the starting grayscale level is usually $1 \times 10^{19} \rm{cm}^{-2}$, it is different for some sources to enhance the grayscale contrast.}
(See Fig. \ref{fig:examples} for two sample plots; the full set is available in the electronic edition of this Journal). 

\begin{table*}[tb!]
\caption{\label{tab:locations} Locations and FUV fluxes of candidate PDRs. The full table is available in the electronic edition of this Journal. }
\centering
\begin{tabular}{cccccccc}
\hline\hline
  Source No. & R.A. (2000) & DEC (2000) & Radius & $F_{FUV}$ & Aperture & Grayscale start & Increment\\
  &  &  & (kpc) & \UVunits & \arcsec & ($10^{19}~\rm{cm}^{-2}$) & \\
\hline
1      &13 37 01.346       & -29 52 44.72 & 1.1 & 5.16  & 19 &1& 1.35\\
2      &13 36 56.700       & -29 51 54.00 & 1.3 & 2.08  & 16 &1& 1.33\\
3      &13 37 03.402       & -29 52 51.43 & 1.5 & 2.20  & 14 &1& 1.37\\
4      &13 36 55.827       & -29 52 13.58 & 1.5 & 2.56  & 18 &1& 1.30\\
5      &13 36 59.158       & -29 53 02.08 & 1.6 & 6.83  & 21 &1& 1.36\\
6      &13 37 02.955       & -29 50 48.84 & 1.6 & 12.24 & 14 &1& 1.37\\
7      &13 37 04.572       & -29 50 58.09 & 1.6 & 6.89  & 12 &1& 1.33\\
8      &13 37 02.445       & -29 53 05.43 & 1.7 & 1.18  & 10 &1& 1.35\\
9      &13 37 08.120       & -29 51 02.97 & 2.4 & 20.81 & 20 &1& 1.37\\
10     &13 36 54.317       & -29 53 10.26 & 2.5 & 21.61 & 17 &1& 1.35\\
\hline
\end{tabular}
\end{table*}

The disk of M83 is warped at larger galactocentric radii. This influences the deprojected galactocentric radius that we used to determine the dust-to-gas ratio. The distance from the FUV sources to the associated \HI\ patches is also affected, since we use the position angle and declination to deproject the separation $\rho_{\HI}$ (see \citet{2008ApJ...673..798H} for a more detailed description of the procedure and projection issues). Table \ref{tab:tilted} shows the values we adopted from the tilted ring model of \citet{1974ApJ...193..309R}. 

\begin{figure}[tb!]
  \resizebox{\hsize}{!}{\includegraphics{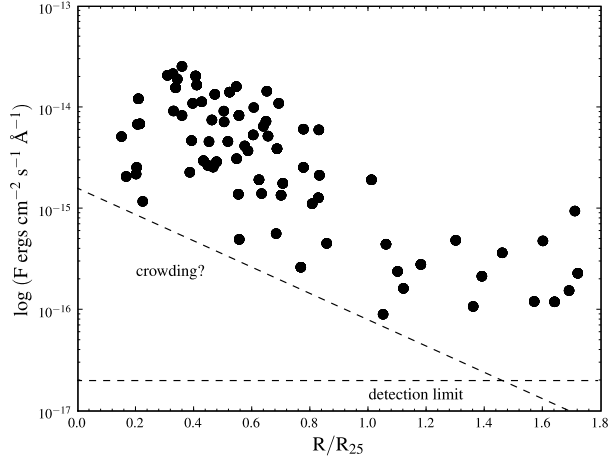}}
  \caption{The FUV fluxes of the candidate PDRs are plotted here on a logarithmic scale. Outside of $R_{25}$ the source fluxes seem to flatten out. A detection limit has been included, as well as a possible crowding limit. The angular resolution of the data is approximately 8\arcsec.}
  \label{fig:F}
\end{figure}

The background subtracted FUV flux of each source is plotted in Fig. \ref{fig:F} and listed in Table \ref{tab:locations}. The background UV level was determined at the radius listed in the \correction{a}perture column. A range of fluxes can be seen, which seems to be decreasing towards the outer parts of M83. However, crowding effects are a possible cause of this trend. \correction{This is indicated in Fig. \ref{fig:F}}. In addition we have made no attempt to obtain a statistically complete sample of sources. A detection limit is plotted based on the noise level of the FUV data and the method we used to select our sources. These fluxes have not been corrected for extinction. We did not select sources close to the central starburst (See e.g. \citet{2005ApJ...619L..83B} for a description of the UV radial profiles found by GALEX).

\begin{figure}[tb!]
  \resizebox{\hsize}{!}{\includegraphics{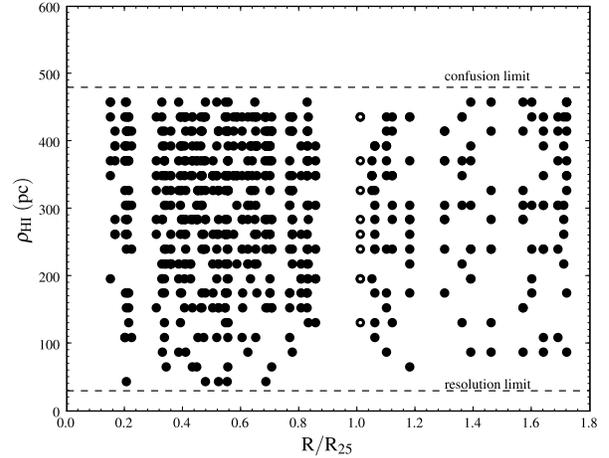}}
  \caption{The distances $\rho_{\HI}$ from the central UV source to individual \HI\ patches are plotted here. Each source can have multiple \HI\ patches associated with it. As an example, 10 \HI\ patches associated with one particular UV source are plotted as open circles here, at $R/R_{25} \sim 1$ (Sources 61 a-j in Table \ref{tab:results}; three sources have an identical separation). At a separation larger than 22\arcsec\ no \HI\ patches are recorded anymore. In the outer parts fewer \HI\ patches are detected, possibly because of sensitivity issues. Some gaps in the radial coverage can be seen, where no sources were found with our selection method.}
  \label{fig:rho}
\end{figure}

In an attempt to capture as much information as possible per candidate PDR, we determined the distance $\rho_{\HI}$ of every \HI\ patch to its central UV source (Fig. \ref{fig:rho}). This means that every FUV source has a number of measurements associated with it, which appear in the plot as a number of points at the same galactocentric radius. At an arbitrary distance of about 480 pc from the FUV source (roughly 22\arcsec) we stopped recording \HI\ patches, since at that distance the association of these patches to that particular UV source becomes highly unlikely. 
\correction{This cut-off is arbitrary, since it becomes progressively less likely for an \HI\ patch to be associated with the central UV source. The cut-off was chosen with the typical size of large scale PDRs in mind (a few hundred parsec) and is labeled 'confusion limit' in the plot. Large scale PDRs are expected to span a few hundred parsec at most and other UV associations will also be encountered. The separations are quantized since we used 1\arcsec\ rings.} In certain cases the candidate PDRs themselves are fairly close to each other. In such a case an \HI\ patch can be attributed to two UV sources, but without further morphological information no distinction can be made. The minimum observable distance is determined by the resolution of our data and is about 30 pc. This plot also shows the coverage in galactocentric radius, which shows a few gaps where no FUV source was found (or selected) using the selection method described earlier. The full results are tabulated in Table \ref{tab:results}, where the different \HI\ patches per FUV source are labeled a, b, c, and so on. $\rho_{\HI}$, $N_{HI}$, the incident flux \Gnaught, the source contrast $G/G_{bg}$, the total hydrogen volume density $n$, and the fractional errors are listed in this table. 

In rare cases the \HI\ column density is very low, where the measured column is close to the value of the subtracted background. No columns higher than $\approx 2.5 \times 10^{21}~\rm{cm}^{-2}$ were found. This is potentially a beam smoothing effect, where we only resolve \HI\ patches partially. We conclude this from our work on M33 (Heiner et al., in prep.). At higher linear resolution we observe higher \HI\ columns, which points to better resolution of the \HI\ clouds in M33.

\begin{table}[ht]
\caption{\label{tab:tilted} Adopted tilted ring values from \citet{1974ApJ...193..309R}.}
\centering
\begin{tabular}{ccc}
\hline\hline
Radius (\arcmin) & $p.a.$ (degrees) $i$ (degrees) \\
\hline
0.5&45&24\\
1.0&45&24\\
1.5&45&24\\
2.0&45&24\\
2.5&46&24\\
3.0&46&24\\
3.5&46&24\\
4.0&46&25\\
4.5&46&26\\
5.0&49&27\\
5.5&52&28\\
6.0&59&30\\
6.5&67&32\\
7.0&75&34\\
7.5&83&36\\
8.0&91&37\\
8.5&99&39\\
9.0&106&40\\
9.5&113&42\\
10.0&120&43\\
10.5&127&45\\
11.0&133&46\\
11.5&139&47\\
12&145&49\\
12.5&150&50\\
\hline
\end{tabular}
\end{table}

\begin{figure}[tb!]
  \resizebox{\hsize}{!}{\includegraphics{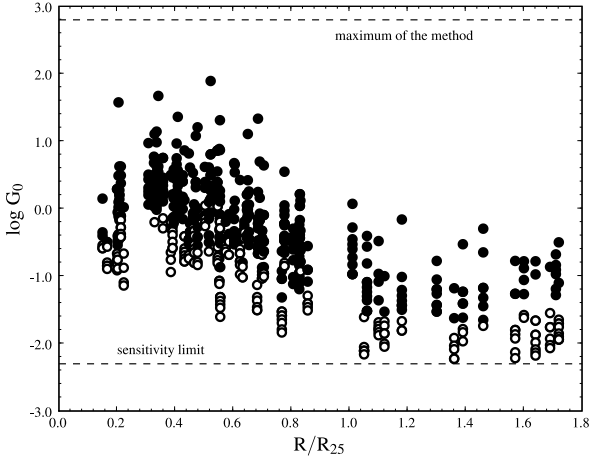}}
  \caption{The incident flux \Gnaught\ is plotted here. Some holes in the radial coverage are visible. The fluxes within $0.2 R_{25}$ seem to be lower. Generally, fluxes are dropping going outward. The maximum of the method consists of the \Gnaught\ that would have resulted from our maximum measured FUV flux, combined with the lowest separation that can be resolved at the given resolution. Sources with a source contrast of less than 0.5 are plotted as open circles.}
  \label{fig:G0}
\end{figure}

The incident flux \Gnaught\ (corrected for foreground extinction) on each individual \HI\ patch is plotted in Fig. \ref{fig:G0}. A set of points in the vertical direction shows all resulting \Gnaught\ for an individual UV source with multiple \HI\ patches. The average \Gnaught\ declines going outward, with no distinct break at the edge of M83's optical disk. A hint of an upturn at the location of M83's \HI\ ring is visible at the extreme right end of the plot. As \Gnaught\ is a combination of the FUV flux and the FUV source / \HI\ patch separation, it is also influenced by the possible crowding effects that are indicated in Fig. \ref{fig:F}.
The minimum \Gnaught\ (0.005) is due to the maximum $\rho_{\HI}$ that we measure, combined with the \correction{lowest} measured flux, and is therefore a sensitivity limit. The highest \Gnaught\ (565) that we could obtain is determined by the \correction{lowest} measurable $\rho_{\HI}$ and the highest FUV flux. Both limits are plotted. The open circles are incident fluxes with a source contrast of less than 0.5. This means that the incident source flux has 50\% of the strength of the ambient FUV field incident on the \HI\ patch or less. The \HI\ corresponding to the closed circles can be assumed to have a higher chance of having been produced by photodissociating UV radiation, since the central UV source is dominant with respect to the background radiation field (the source contrast is higher). However, these highest source contrasts occur when the separation between the central UV source and the associated \HI\ patch is the \correction{lowest} - at close separation projection effects are more of a problem, which means a higher chance of underestimating the separation and overestimating the incident flux. A high source contrast at an \HI\ patch therefore needs to be viewed with caution.

\begin{figure}[tb!]
  \resizebox{\hsize}{!}{\includegraphics{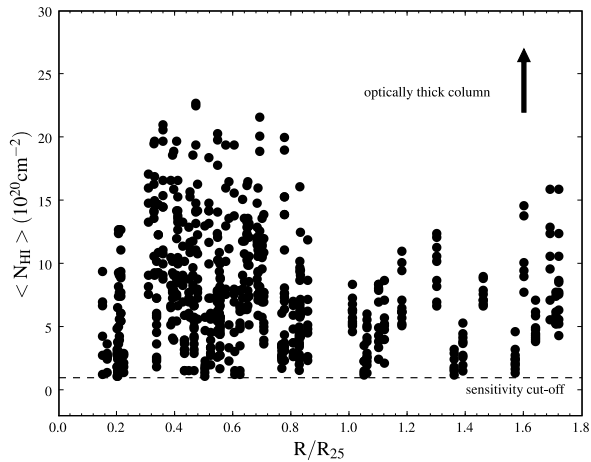}}
  \caption{The (beam averaged) \HI\ column densities are plotted here. Values range from around the sensitivity limit at $1 \times 10^{20}~\rm{cm}^{-2}$ to almost $2.5 \times 10^{21}~\rm{cm}^{-2}$. Columns are highest within $R_{25}$ and a small upturn can be seen signifying M83's \HI\ ring at about 1.7 $R_{25}$.}
  \label{fig:HI}
\end{figure}

The (beam averaged) atomic hydrogen column density of each \HI\ patch in our candidate PDRs is plotted in Fig. \ref{fig:HI}. A selection of local maxima are counted as \HI\ patches. Near the center of M83 there is little \HI\ emission visible against the continuum source at the nucleus \citep[e.g.][]{1993A&A...274..707T}.
The maximum observed column density rises to a broad peak around 0.5 $R_{25}$. The end of M83's disk is clearly visible where the maximum \HI\ column densities drop, but \HI\ is still detected. Finally, there is a slight increase where M83's \HI\ ring is located.

An \HI\ background value was subtracted as described previously, in an attempt to capture only the \HI\ produced by the incident FUV radiation from the FUV source nearby. \correction{The resulting values of $n$ are relatively insensitive to varying the background \HI\ column.}
Background-subtracted values below the sensitivity limit were discarded.

\begin{figure}[tb!]
  \resizebox{\hsize}{!}{\includegraphics{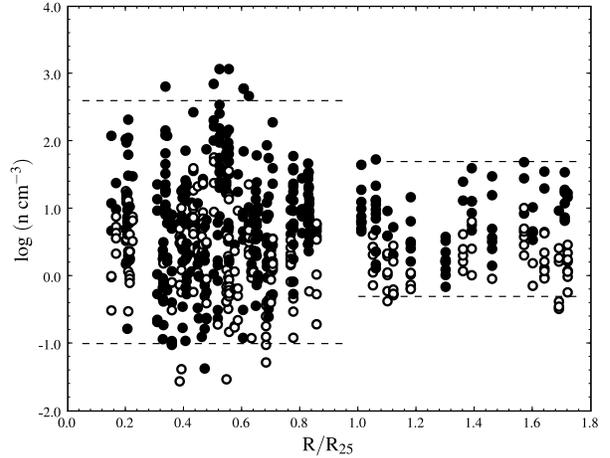}}
\caption{The total hydrogen volume density is plotted, assuming relatively high metallicity (our preferred dust model, Equation \ref{eqn:GP07h}) outside the optical disk. Closed circles indicate a source contrast \correction{higher} than 0.5, open circles indicate a lower source contrast. An approximate range of values of $n$ is indicated (dashed lines): 0.1 - 400 cm$^{-3}$ inside $R_{25}$, 0.5 - 50 cm$^{-3}$ outside $R_{25}$.}
  \label{fig:Ht}
\end{figure}

The calculated total hydrogen volume density is shown in Fig. \ref{fig:Ht}, based on the \citet{2007ApJ...661..115G} high metallicity assumption, which is our preferred dust model. Values span a range of two orders of magnitude mostly, with values of a few thousand $\rm{cm}^{-3}$ in the inner parts of M83. Outside M83's optical disk the values of $n$ are generally lower. In this plot, again closed circles are plotted for \HI\ patches that show a source contrast \correction{higher} than 0.5 (meaning the source is half as bright as the UV background or stronger), and open circles for those with lower source contrast. The former may be considered more reliable than the latter, since the UV source is considered to be more clearly responsible for the dissociated \HI. The FUV source contrasts $G/G_{bg}$ are listed in Table \ref{tab:results}.  The fractional errors in Table \ref{tab:results} are as calculated in the \citet{2008ApJ...673..798H} M81 results, except that we used a relative error for the dust-to-gas ratio of 20\% instead of a constant error. The average fractional error is 0.7. Since photodissociation rates are highly sensitive to the local dust-to-gas ratios, we looked at a number of recent results for M83.

\begin{figure}[tb!]
  \centering
  \includegraphics[width=8.5cm]{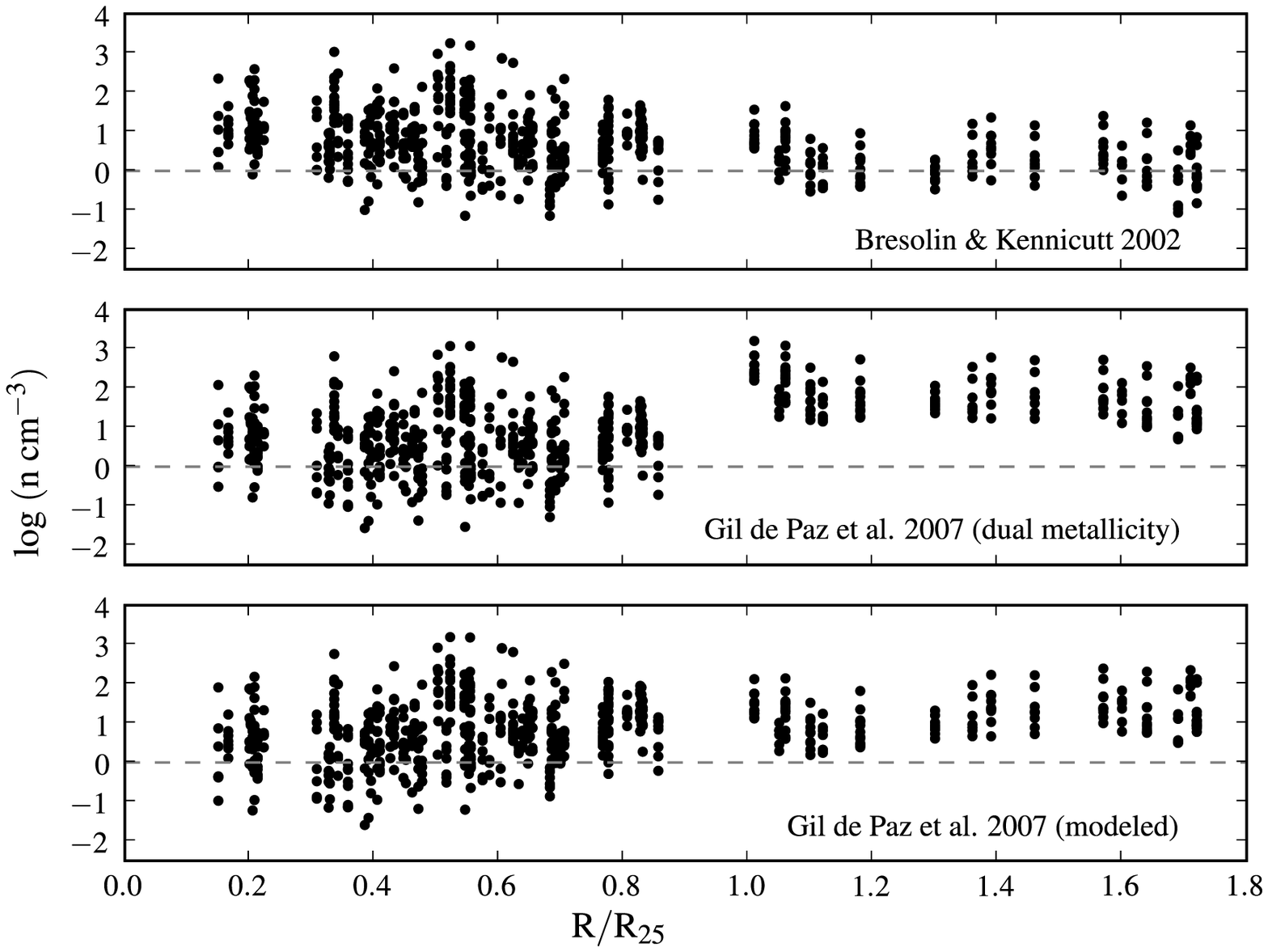}
  \caption{The total hydrogen volume density is plotted for three dust models derived from \citet{2002ApJ...572..838B} and \citet{2007ApJ...661..115G}. The results based on the \citet{2002ApJ...572..838B} dust model are similar to the one from \citet{2007ApJ...661..115G} assuming a high metallicity outside the optical disk (Fig. \ref{fig:Ht}). In the dual metallicity case, the hydrogen densities are relatively high in the outer regions. The single slope fit based on PDR modeling by \citet{2007ApJ...661..115G} is steep because of relatively high metallicities near the center of M83 and low metallicities in the outer regions. Densities are markedly lower in the center in this case.}
  \label{fig:Htcombo}
\end{figure}

Different alternative total hydrogen volume densities based on various dust models suggested by \citet{2002ApJ...572..838B} and \citet{2007ApJ...661..115G} are shown in Fig. \ref{fig:Htcombo}. The different dust models can be found in Fig. \ref{fig:dust}. The metallicity model based on \citet{2002ApJ...572..838B} uses
\begin{equation}
  \log{\delta/\delta_0} = -0.021 R + 0.17.
  \label{eqn:BK02}
\end{equation}
The slope of this model is relatively shallow. The dual metallicity model in the middle panel uses Equation \ref{eqn:GP07h}, except that $\delta/\delta_0 = 0.15$ for all PDRs outside of $R_{25}$, assuming low and constant metallicities in the outer regions of M83 (Gil de Paz, private communication, 2008). The bottom panel uses
\begin{equation}
  \log{\delta/\delta_0} = -0.100 R + 0.85.
  \label{eqn:GP07}
\end{equation}
The slope of the metallicity is relatively steep in this case since a single slope is assumed, with high metallicities inside the optical disk of M83 and low metallicities outside.

The top panel densities are similar to the ones in Fig. \ref{fig:Ht}, since the underlying dust models are also very similar. The dual metallicity model in the middle panel yields densities in the outer regions that appear to be unusually high. Finally, the densities in the bottom panel are extremely depressed at small galactocentric radii, a result of the high dust content. Insufficient data are available to pick one dust model over the other. We continue our analysis using the high metallicity model of \citet{2007ApJ...661..115G}, as shown in Fig. \ref{fig:Ht}.

\begin{figure}[tb!]
  \resizebox{\hsize}{!}{\includegraphics{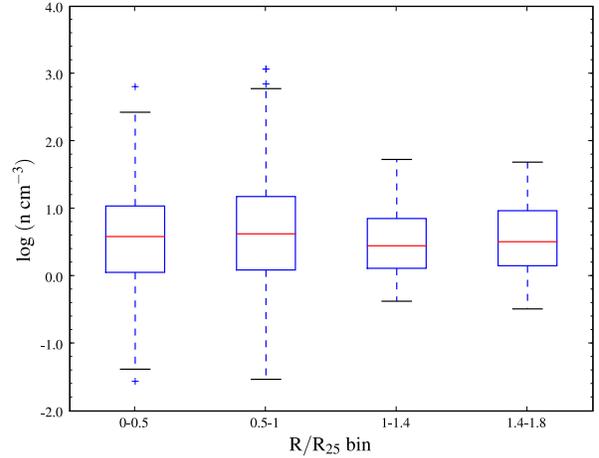}}
  \caption{Box-and-whisker plots for the total hydrogen volume densities in Fig. \ref{fig:Ht} are displayed, where the data has been divided into four galactocentric radius bins. The boxes span the lower to upper quartiles of the data (50\% of the datapoints in the bin combined), with a line indicating the median. The so-called 'whiskers' show the range of the data (1.5 times the inner quartile range of points), with outliers plotted individually (+). The approximate ranges in Fig. \ref{fig:Ht} are in good agreement with this. The maximum densities in the outer regions are lower than the ones in the inner regions of M83 and the range of values is \correction{narrower}.}
  \label{fig:Htboxplot}
\end{figure}

In order to further quantify the volume densities in Fig. \ref{fig:Ht}, so-called box-and-whisker plots are shown in Fig. \ref{fig:Htboxplot}. Since we did not aim to select a complete sample of candidate PDRs, we are mostly interested in the range of volume densities. The galactocentric radius bins in this plot give a good impression of the distribution of the volume densities. There is a \correction{wider} spread of values in the inner regions of M83, accompanied by \correction{higher} density values. The values in the outer regions are more concentrated and the maximum densities are lower. This is mostly due to the \correction{narrower} spread of \HI\ column densities outside $R_{25}$. Selection effects also limit the range of observed densities.

\begin{figure}[tb!]
  \resizebox{\hsize}{!}{\includegraphics{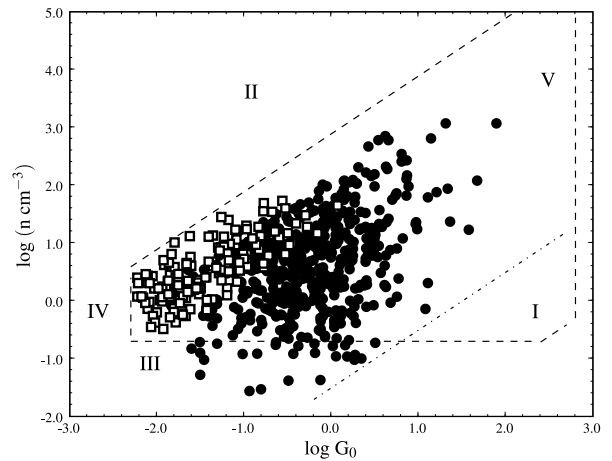}}
  \caption{Our results can be summarized in a plot of \Gnaught\ against $n$. The numerals indicate observational selection effects discussed in the text. In this plot, the open squares indicate regions outside of $R_{25}$.}
  \label{fig:G0n}
\end{figure}

Our results are summarized in Fig. \ref{fig:G0n}, which is suitable to show the various limiting selection effects. The candidate PDRs at a galactocentric radius larger than $R_{25}$ are plotted as open squares. 
The Roman numerals I through V indicate the different limits of our observations:

\makeatletter
\renewcommand{\theenumi}{\Roman{enumi}}
\renewcommand{\labelenumi}{\theenumi.}
\makeatother
\begin{enumerate}
  \item \correction{The} \HI\ column density upper limit of $5 \times 10^{21}~\rm{cm}^{-2}$ at a typical \dtg\ of 1.8, due to the \HI\ column becoming optically thick. However, the actual upper limit of the \HI\ is closer to $2.5 \times 10^{21}~\rm{cm}^{-2}$ because of beam smoothing effects. These values occur around a typical \dtg\ of 2.4 and are delineated by an extra dashed-dotted line.
  \item \correction{The} \HI\ lower limit of $1 \times 10^{20}~\rm{cm}^{-2}$ at a typical \dtg\ of 1. This is a sensitivity limit.
  \item The lowest volume density that we can observe as a consequence of the radio beam diameter and the \HI\ sensitivity limit. For M83, it is approximately $0.2~\rm{cm}^{-3}$. Points may end up lower because of high metallicities. Ideally, this limit would contain the GMC cloud size. Since this is unknown, the beam diameter is a reasonable substitute.
  \item The minimum \Gnaught\ (0.005) that we can use (at a source contrast as low as 1\%).
  \item The maximum \Gnaught\ (565) that we can obtain from the data.
\end{enumerate}

\begin{table*}[tb!]
\caption{\label{tab:results} \HI\ Measurements, Incident Fluxes and Volume Densities}
\centering
\begin{tabular}{ccccccc}
\hline\hline
  Source No. & $\rho_{\HI}$ (pc) & $N_{\HI}$ (\HIunits) & 
  \Gnaught & $G/G_{bg}$ & $n$ ($\rm{cm}^{-3}$) & Fractional error\\
  \hline
1a  & 196	& 1.26 & 1.41 &  2.05 & 120      &  0.8     \\
1b  & 349	& 2.77 & 0.45 &  0.65 & 12       &  0.7     \\
1c  & 371	& 4.28 & 0.40 &  0.57 & 5        &  0.7     \\
1d  & 436	& 6.99 & 0.29 &  0.41 & 1        &  0.9     \\
1e  & 458	& 6.69 & 0.26 &  0.38 & 1        &  0.9     \\
1f  & 458	& 9.40 & 0.26 &  0.38 & 0.3      &  1.2     \\
2a  & 262	& 1.41 & 0.32 &  0.57 & 24       &  0.7     \\
2b  & 262	& 2.62 & 0.32 &  0.57 & 10       &  0.7     \\
2c  & 284	& 2.77 & 0.27 &  0.49 & 8        &  0.7     \\
2d  & 371	& 2.47 & 0.16 &  0.28 & 5        &  0.7     \\
2e  & 393	& 2.62 & 0.14 &  0.25 & 4        &  0.7     \\
2f  & 393	& 2.92 & 0.14 &  0.25 & 4        &  0.7     \\
2g  & 415	& 3.67 & 0.13 &  0.23 & 2        &  0.7     \\
\hline
\end{tabular}
\end{table*}

\section{Discussion and Conclusions}
\label{sec:discussionconlusions}

In this paragraph we compare our results to \Htwo\ densities based on CO measurements. Then we discuss background issues, extinction and dust issues. We end with a summary of our findings.

\subsection{CO results comparison}
\label{sub:CO}

The study of \citet{1999ApJ...513..720R} is detailed enough to find typical sizes and masses of GMCs in the eastern arm of M83. These results can be compared to ours with some additional assumptions. Using a CO(1-0) map with kinematic information, they find CO masses and virial masses onindicatingindicating the order of $1 \times 10^6 M_{\odot}$. For example, their source 9 (their Table 4) yields a mass derived from CO emission of 7.3 $\times 10^6 M_\odot$ and 3.6 $\times 10^6 M_\odot$ for the virial masses, at a 50-80\% uncertainty. This source corresponds to our FUV source no. 13. Since our method does not yield any information on cloud sizes (we only observe HI on the surface of GMCs), nor on kinetic temperatures in these clouds, we will take typical values of our measured volume densities and compare densities using a typical GMC radius of 75 pc for a spherical cloud. $n = 2 n_{H_2}$ inside the GMC and a typical density near our source no. 13 is $20~\rm{cm}^{-3}$, or $n_{H_2} = 10~\rm{cm^{-3}}$. 7.3 $\times 10^6 M_\odot$ is equivalent to about $n_{H_2} = 85 \rm{cm^{-3}}$, which is in reasonable agreement with our results considering the uncertainties in both results (70-80\%). More recent unpublished data (Lord, private communication, 2008) may be quantitatively different and more detailed, but does not yield substantially different results at the resolution of the data that were used in this paper.

One of the key differences between M81 and M83 is the abundance of CO emission. While CO is extremely faint and hard to detect in M81 \citep[e.g.][]{2007A&A...473..771C}, M83 displays bright CO features. The method we used here to find hydrogen densities (and molecular hydrogen densities in GMCs) does not yield any morphological information, since we are observing \HI\ on the surface of PDRs. We can note, however, that the size and scale we assume for our candidate PDRs (the range of values of $\rho_{\HI}$) is consistent with findings using CO. Since we did not find suitable candidate PDRs in the inner kiloparsec of M83, we cannot compare our results with for example the results of \citet{1993ApJ...414...98W}, although the densities we find match their low-density component. As to the truncation of the molecular disk that was mentioned by \citet{2002AJ....123.1892C}, we do see a difference between the inner and outer disk of M83, but it does not seem to be that significant in our results. No differences are seen in candidate PDRs in arm or inter-arm regions, insofar as the large scatter in our results allows us to draw this conclusion. We do not find more than a handful of regions with the densities indicated by \citet{2007ApJ...664..363H}, although the expected sizes of our candidate PDRs are the same. Finally, the hydrogen densities presented here are similar to the M81 results but extend to higher maximum values. This is consistent with the brighter CO emission in M83, since the higher \Htwo\ densities will lead to a greater degree of excitation of the CO molecules. Alternatively, this could also indicate a higher fraction of CO molecules in M83's clouds \citep[e.g.][]{2007A&A...473..771C}.

\subsection{Background levels, extinction and dust}
\label{sub:FUV}

The FUV fluxes seem to be higher in the inner parts of M83 and decreasing (on average) going outward, as is shown in Fig. \ref{fig:F}. These higher fluxes in the inner parts could be caused by brighter sources, and/or a larger number of sources. Since the individual sources are unresolved, no firm conclusions can be drawn here. Outside $R_{25}$ the measured fluxes are roughly constant. At the same time, the abundances of PDR-produced \HI\ surrounding these FUV sources (Fig. \ref{fig:HI} - note that these are individual measurements, not annular averages) seem to follow a similar trend, consistent with its connection to the photodissociating UV radiation. The FUV source contrast does not vary with galactocentric radius, so while both source and background radiation field decrease in intensity towards the outer regions of M83, their relative strength does not change.

Another factor that could influence the FUV radiation is galactic foreground extinction. The higher the applied foreground extinction, the higher the resulting total hydrogen volume density. The 0.52 mag extinction \citep{1998ApJ...500..525S} towards M83 is significantly less than the 1.37 mag we used towards M81 in \citet{2008ApJ...673..798H}. The \citet{1998ApJ...500..525S} extinction correction for M81 would be 0.58, significantly lowering the total hydrogen volume densities. We find a \correction{wider} range of values in M83, including higher gas densities. We therefore expect M83 to harbor more gas than M81. The M81 and M83 results are hard to compare directly because the extinction corrections were based on different sources in the literature. We intend to compare the two galaxies, together with similar M33 results, more quantitatively in a future paper, with consistent extinction corrections.

Our results are also sensitive to the dust-to-gas ratio. The slope of M83's metallicity is relatively shallow. Our preferred dust model assumes that metallicities in the outer parts of M83 remain high. Lower metallicities would result in higher gas densities. Our results in these regions are therefore most likely lower limits and much more gas could be present. The recent results by \citet{2007ApJ...661..115G} are ambiguous in the sense that the authors provide high metallicity and low metallicity results, depending on the adopted model. In the high metallicity scenario, the slope of M83's metallicity remains shallow out to large galactocentric radius. The low metallicity scenario is accompanied by a sharp drop in metallicity starting at approximately $R_{25}$.

\citet{2000ApJ...538..608S} found a similar range of gas densities in M101 using basically the same method with additional extinction corrections. Their candidate PDRs were all within M101's $R_{25}$, but the FUV luminosities within that range are similar.
As in our previous M81 results, no internal extinction correction was applied. If any such correction were applied, it should scale with the dust-to-gas ratio. This means that the extinction would decrease going outward. The FUV fluxes would be even higher in the center of M83 (leading to higher total hydrogen volume densities), while the fluxes in the outer regions of M83 would not be affected much. 

\subsection{Summary}
\label{sub:summary}

In summary, we have investigated atomic hydrogen found in candidate PDRs in M83 and used the physics of these PDRs to derive total hydrogen volume densities. We carefully considered the contribution of observational selection effects to our results. We find a range of densities: 0.1 - 400 cm$^{-3}$ inside $R_{25}$, 0.5 - 50 cm$^{-3}$ outside $R_{25}$, based on measurements of \HI\ believed to be produced in large scale PDRs in M83. The higher GMC volume densities which we find within $R_{25}$ correlate well with the presence of bright CO(1-0) emission there. This points to enhanced collisional excitation as one reason for the CO emission in these GMCs. We note that this is also consistent with our results in M81, where we find little evidence for high GMC volume densities, and for which the CO(1-0) emission is faint.

Our measurements go out to a galactocentric radius of 13 kpc (deprojected). Our study used the tilted ring model from \citet{1974ApJ...193..309R} to get proper galactocentric radii outside the M83 optical disk. Our results are notably sensitive to the local dust-to-gas ratio, especially since the metallicity in the outer regions remains uncertain.

\begin{acknowledgements}
  The authors are grateful to Fabian Walter, Elias Brinks and the THINGS team for providing us with M83 radio data as well as for engaging in useful discussions. We would also like to thank Gerhardt Meurer for his help and useful discussions, as well as David Thilker, Armando Gil de Paz and Steven Lord. \correction{We thank the anonymous referee for providing suggestions that improved the clarity of this work.} J.S.H. acknowledges the support of a Graduate Research Assistantship provided by the STScI Director's Discretionary Research Fund.
\end{acknowledgements}

\bibliographystyle{aa}
\bibliography{references}

\clearpage

\end{document}